# Dependence of technological improvement on artifact interactions

Subarna Basnet, Christopher L. Magee
Massachusetts Institute of Technology, SUTD-MIT International Design, 77 Mass Ave, Cambridge, MA-02139

## Abstract

Empirical research has shown performance improvement of many different technological domains occurs exponentially but with widely varying improvement rates. What causes some technologies to improve faster than others do? Previous quantitative modeling research has identified artifact interactions, where a design change in one component influences others, as an important determinant of improvement rates. The models predict that improvement rate for a domain is proportional to the inverse of the domain's interaction parameter. However, no empirical research has previously studied and tested the dependence of improvement rates on artifact interactions. A challenge to testing the dependence is that any method for measuring interactions has to be applicable to a wide variety of technologies. Here we propose a patent-based method that is both technology domain-agnostic and less costly than alternative methods. We use textual content from patent sets in 27 domains to find the influence of interactions on improvement rates. Qualitative analysis identified six specific keywords that signal artifact interactions. Patent sets from each domain were then examined to determine the total count of these 6 keywords in each domain, giving an estimate of artifact interactions in each domain. It is found that improvement rates are positively correlated with the inverse of the total count of keywords with correlation coefficient of +0.56 with a p-value of 0.002. The empirical results agree with model's predictions and support the suggestion that domains with higher number of artifacts interactions (higher complexity) will improve at a slower pace.

## Statement of significance

Qualitative and quantitative research has suggested that artifact interactions influence technological performance improvement. The current study provides, to our knowledge, the first empirical evidence that artifact interactions contribute to variation in improvement rates. The study shows artifact interactions including component-to-component or component-to-system interactions, coupling of functional requirements, and concomitant side effects in artifacts have a retarding effect on performance improvement, thus implying that internal complexity of artifacts plays a vital role in constraining the evolution of performance over time. The study also for the first time demonstrates that patents can be a useful and attractive resource for studying interactions as our empirical measure of interactions in domains were derived using textual content from patents over four decades.

## Introduction

Within the large and complex field of technical change, empirical research has demonstrated technological performance improves exponentially over time, but with widely varying improvement rates across the domains. In addition, knowledge of how and at what rate performance of a given technology improves is important for corporate product planners and designers, policy makers, and investors (1, 2, 3, 4, 5, 6, 7). To improve consistency and reduce ambiguity in measurement of technology performance and its improvement, Magee et al. (7) have chosen technological domains as the unit of analysis, which they define as a set of designed artifacts that utilize a recognized body of knowledge to achieve a specific generic function. The artifacts considered can be products, software, or processes. The body of knowledge is principally scientific and engineering knowledge of particular relevance to the domain of interest; so each functional category is decomposed into technological domains based on the scientific knowledge utilized by the artifacts considered. The performance metric of a technological domain, defined from the perspective of users of technology, is a composite indicator which includes essential functional outputs and a resource constraint (e.g., cost, volume or mass of the artifact) important to the users. The performance metric is formulated so that as the functional outputs improve the performance metric increases, and is expressed per unit of resource considered. The available data has been, accordingly, adapted to construct performance data for 71 metrics in 27 domains (7). The analysis of empirical data has demonstrated that performances of the 27 technological domains considered improve exponentially, but the annual improvement rates ($K_J$) vary widely ranging from 3 to 65 percent (shown in Fig. 1 for the most reliable estimates).

A quantitative model of the effect of interactions on technological improvement was developed first for simple artifacts and changes in cost (8). For a simplified artifact with interaction parameter d for each component (defined as an average number of components affected by a design change in a given component, including itself), McNerney et al.'s treatment (8) for unit cost results in the following relationship:

$dC/dm = - B \cdot C^{d+1}$ (1)

Where, C = unit cost normalized with respect to initial cost*, m = number of attempts, d= interaction parameter, B = constant

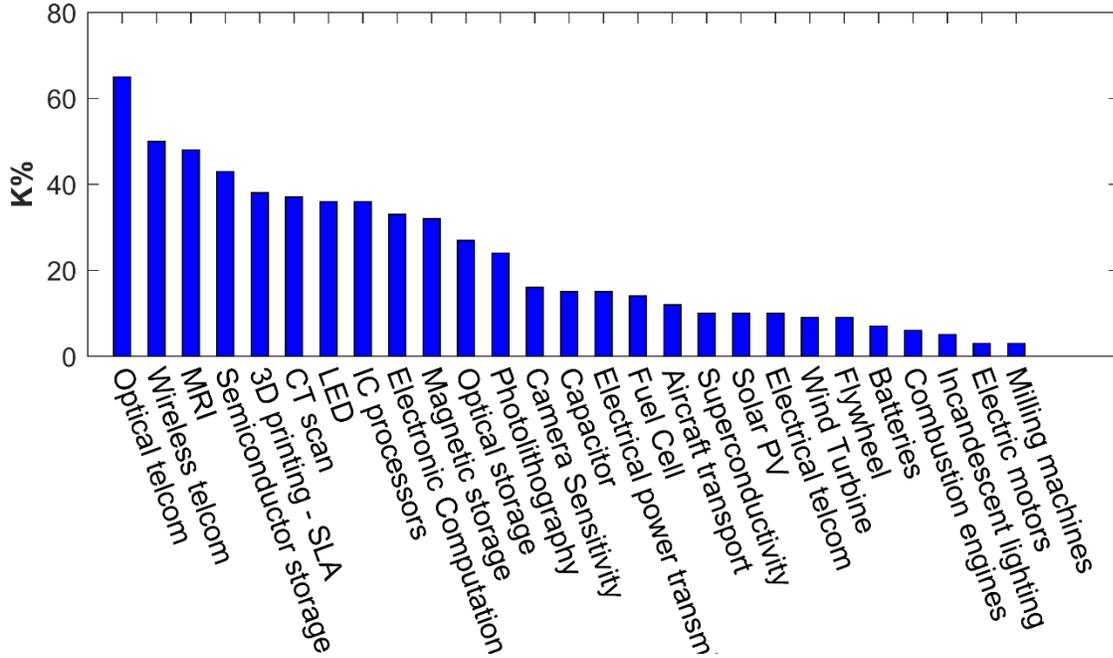

Fig. 1 Annual performance improvement rates ($K_J$) for 27 domains. The value of $K_J$ for each domain is the slope of a linear curve fitted to log of performance metric versus time using data from 1976 until 2013. The rates vary widely, from 3 to 65 percent. Adapted from Magee et al. 2016.

Considering a broader framework, Basnet and Magee (9) and Basnet (10) have extended these results into the form tested here. Their overall result is summarized by an equation giving improvement rate of a technological domain based on three factors on the right hand side of equation 2:

$$K_J = \frac{d \ln Q_J}{dt} = (\pm)A_J \frac{1}{d_J} K \qquad (2)$$

Where $Q_J$ is the performance of a domain J and $K_J$ its performance improvement rate (or, the slope of the performance curve in a semi-log plot with time); $A_J$, and $d_J$ are respectively domain specific scaling and interaction parameters; K is domain independent, and represents the rate of growth in the number of individual operating ideas (IOI) generated through combinatorial analogical transfer process. These IOI are then assimilated into the components of domain artifacts, where the influence of interaction and scaling parameters manifest. The equation predicts that $K_J$ (improvement rate) is proportional to the inverse of interaction parameter $d_J$, implying that domains with higher levels of interactions ($d_J$) will experience slower improvement rate. The goal of this empirical study is to test this dependence of improvement rates on artifact interactions. So, what are artifact interactions?

---

* The normalized unit cost is 1 or less so increases in d in equation 1 result in less improvement per attempt.

In design of artifacts, Simon (11) introduced the notion of interactions in his essays on the complexity of artifacts. When a design of an artifact is changed from one state to another (with differences between the two states as defined by multiple attributes, say D1, D2, and D3) by taking some actions (say, A1, A2, and A3), in many cases, any specific action taken may affect more than one attribute, thus potentially manifesting as interactions of the attributes. The same notion of interaction/conflicts is captured by the concept of coupling of functional requirements (12), or dependencies between characteristics (13), which can occur when two or more functional requirements are influenced by a common design parameter. Theoretically, it seems ideal to have one design parameter controlling one functional requirement to achieve a fully decomposable (modular) design (12, 14). Using an in-depth qualitative analysis of VLSI (very-large-scale integration) systems and complex electro-mechanical-optical (CEMO) systems, Whitney (15, 16), however, has given detailed arguments that, in reality, the decomposability of a design of an artifact depends on the physics involved and/or additional design or resource constraints, such as permissible mass. These couplings or constraints manifest as component-to-component, and component-to-system interactions, or as a need to have multi-functional components. Whitney noted that CEMO artifacts operate by processing significant amounts of power while VLSI artifacts operate by processing information, usually in electrical form, at very low levels of power. Whitney showed that designers face fundamentally different challenges designing these two types of artifacts. The efficiency of CEMO artifacts depends on matching input and output impedances, which inherently couples the components, whereas VLSI designs decouple the components by deliberately, hugely mismatching the impedances. Furthermore, high power levels create side effects at similar power levels, forcing CEMO designers to spend a large portion of their effort predicting and mitigating these side effects. CEMO systems additionally face several resource constraints on mass, size, and power consumption, forcing their designs to conserve these resources by utilizing multi-function components, further increasing inter- and intra-component interactions. During design of new, improved artifacts, the presence, and thus the resolution, of these different interactions causes significant delay, diverts significant engineering resources and potentially stops applications of some concepts, thus making the level of interactions of a technological domain a potentially strong factor influencing its rate of improvement.

The current research utilizes textual content from a set of domain patents to study artifacts interactions. Specifically, we utilize a qualitative approach to identify text describing artifact interactions using keywords as markers. The method (*Material and Methods*) consists of three broad steps: (1) preparing text from the 100-most cited patents in each of the 27 domains (2) determining key words that generically signal instances of interactions in selected patents (3) text mining the text from a patent set for these keywords and arriving at an estimate of the relative number of interactions in the domains (4) testing the prediction (from the models) that improvement rates are positively correlated with the inverse of a normalized count of keywords: $K_J \alpha\ 1/$ count of keywords.

## Results and analysis

**Total count of words across domains:** The total count of words (all text including keywords) in patent text varies widely between the patents (*SI*), with the ratio of total word count between the patent with the highest word count to patent with the lowest count was more than 100. However, when the total word count is compared at the domain level, the distribution of total word count is much narrower, shown in Fig. S1 in *SI*. The domain level total word count ranges from roughly 191,000 down to 95,000, a ratio of slightly over two. The domain with the five highest total word count in descending order are genome sequencing[†], 3D printing, optical memory, CT scan, wireless telecommunications. The domains with lowest count in ascending order are electric motor, electrical telecommunications, milling machine, optical telecom, and flywheel. Since domains with more text can potentially have higher occurrence of keywords, the variation in total word count between domain patent sets indicates that it is necessary to normalize keyword count with respect to total word count.

**Normalized count of keywords (KW):** We have identified six keywords - *prevent, undesirable, requirement, fail, disadvantage, and overcome* - as reliable markers of text describing artifact interactions. We do this based on level of occurrence (intensity of signal), cross-domain usage (useful for evaluating different domains) as well relevancy of keywords (true positives). (*See Methods and material for further details.*) The count of 6

---

[†] Reading of genome sequencing patents shows they have much higher amounts of chemical formulas than any other domain. Since keywords representing interactions are not found in these formulas, the normalized count was highly distorted for this domain so it is eliminated in all analysis reported here.

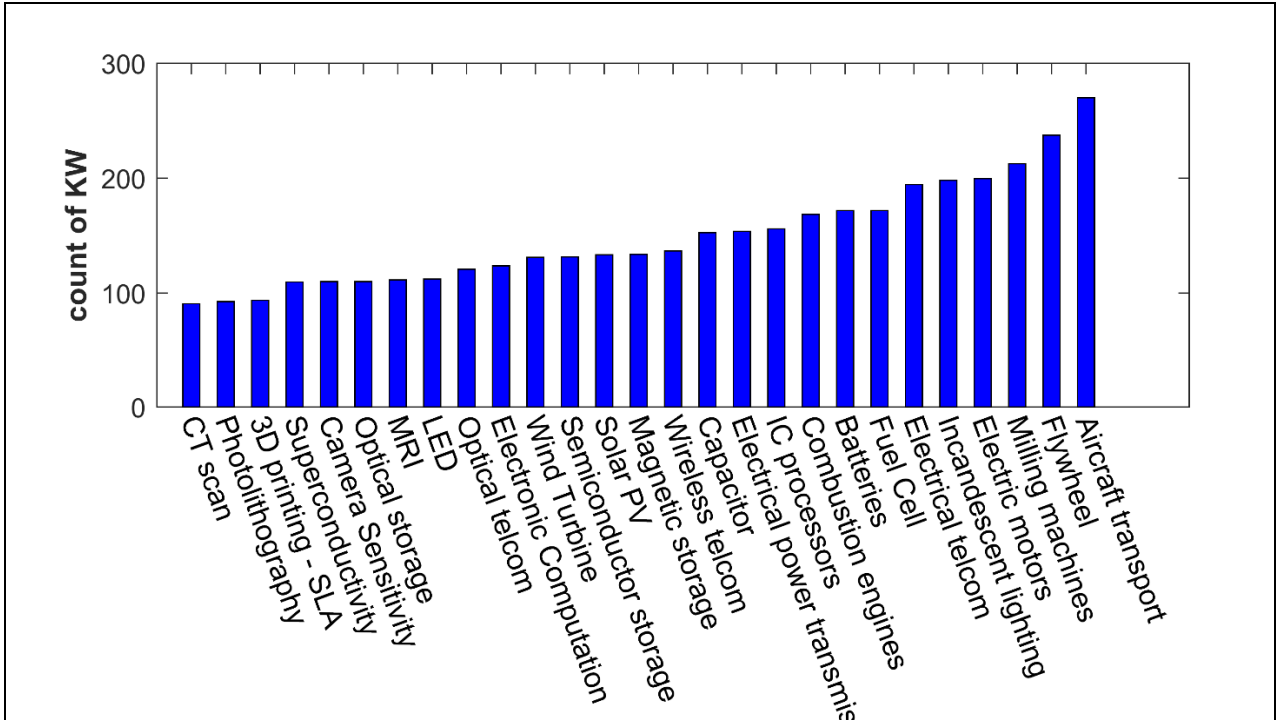

Fig. 2: Comparison of count of normalized 6-keywords (KW) for 27 domains. The count of KW is the normalized total count of 6 keywords identified to indicate interactions in the text (abstract, title, background, and summary of invention) from the 100 most-cited patents in each domain, where normalization is carried out with respect to total number of all words in the text, and expressed per 100 thousand words.

keywords is summed, normalized with respect to total word count in the respective patent set, and finally expressed in terms of keywords per 100,000 words. Fig 2 shows the distribution of normalized count of 6 keywords across the 27 domains. The five domains with the highest normalized 6-keyword count in descending order are Aircraft, Electric power transmission, Flywheel, Electric telecommunication, and Milling machine. The domains with the lowest count in ascending order are CT scan, Superconductors, MRI, 3D Printing-SLA.

**Correlation analysis of improvement rates and reciprocal of normalized count of keywords:** According to the models, the improvement rates should be inversely proportional to the interaction parameter $d_J$. The normalized keyword count is assumed to provide a relative estimate of artifact interactions encountered in generating inventions in the various domains: we assume a linear relationship between the normalized keyword count ($KW_J$) and the number of interactions characteristic of a domain ($d_J$). With this assumption, the prediction in Equation (2) then becomes

$$K_J = \frac{d\,lnQ_J}{dt} \; \propto \; \frac{1}{d_J} \; \propto \; \frac{1}{KW_J} \qquad (3)$$

Relationship 3 is expressed as a hypothesis as follows:

Hypothesis: *The performance improvement rate of technological domains are positively correlated with the inverse of the normalized count of interaction keywords in a set of patents belonging to each of the domains.*

The hypothesis was then tested empirically using the 6-keyword results. The improvement rate in a domain is plotted as a function of the inverse of normalized count of the 6-keywords in the same domain in Fig. 3. Although there is significant scatter in the data, a clear upward trend can be visually observed, implying that higher improvement rates ($K_J$) positively correlate with higher values of the inverse of normalized count of the 6-keywords ($KW_J$). Pearson's correlation coefficient calculated using EXCEL2013 is +0.56 with a p-value of 0.002. The p-value is much smaller than 0.05, a threshold value employed by many researchers indicating a very strong likelihood that the correlation is meaningful. This result supports the theoretical prediction that domains associated with higher number of interactions improve at a slower pace. How reliable is this correlation? We examine the reliability of this finding using a robustness study.

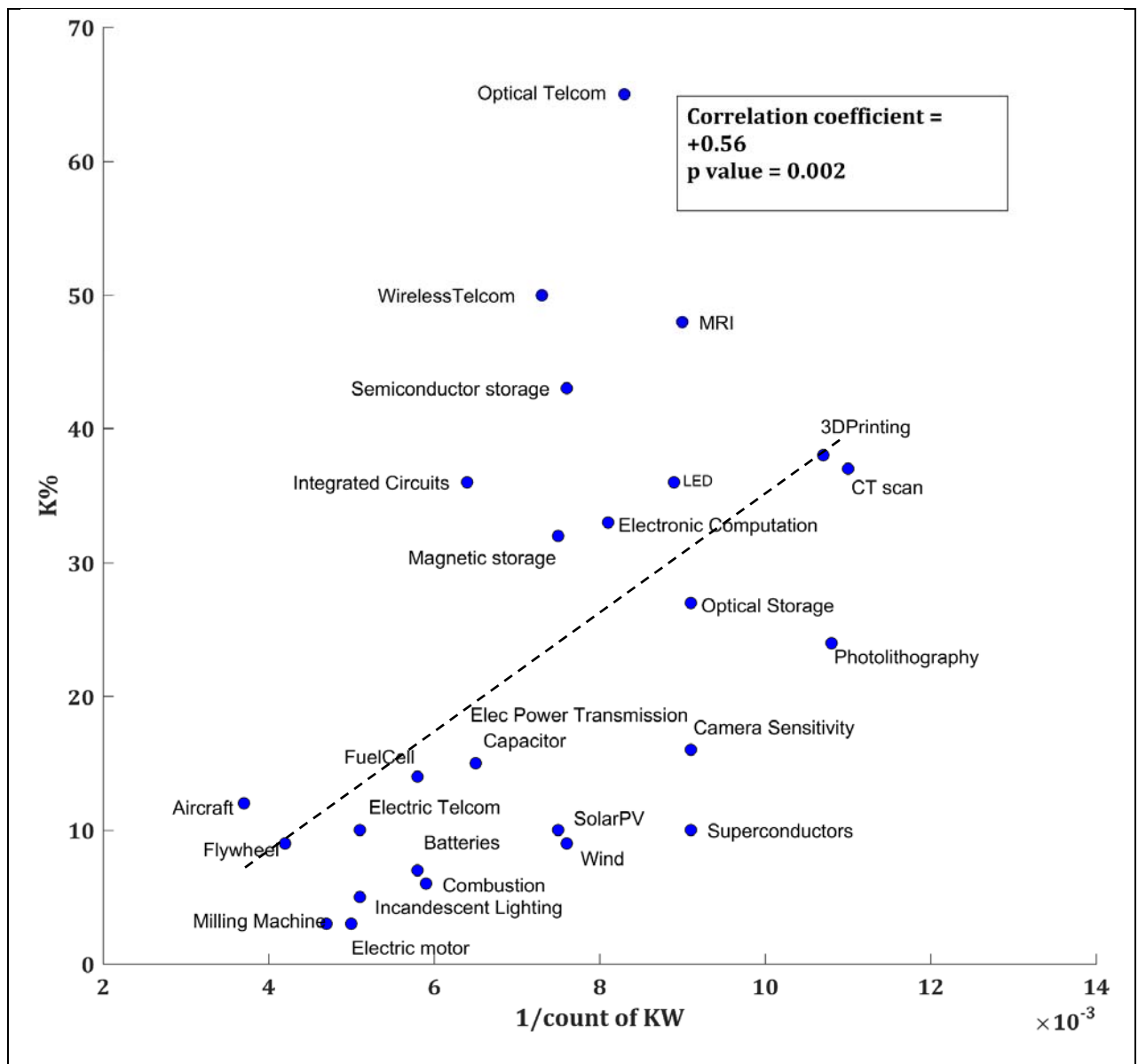

Fig. 3 Scatter plot of $K_J$ (improvement rates) and reciprocal of normalized count of keywords (1/KW) for 27 domains. Improvement rates are positively correlated with Pearson correlation coefficient = +0.56 with p-value = 0.002. Note that capacitor and electric power transmission data points overlap, and hence appear as one point. Dashed line shows a linear trend line between $K_J$ and 1/KW.

**Robustness test:** The robustness study was conducted by creating 20 groups of 14 domains, where each group was generated by randomly selecting a combination of 14 domains from the 27 domains. For each group of 14 domains, the correlation coefficient between K and 1/KW was calculated.

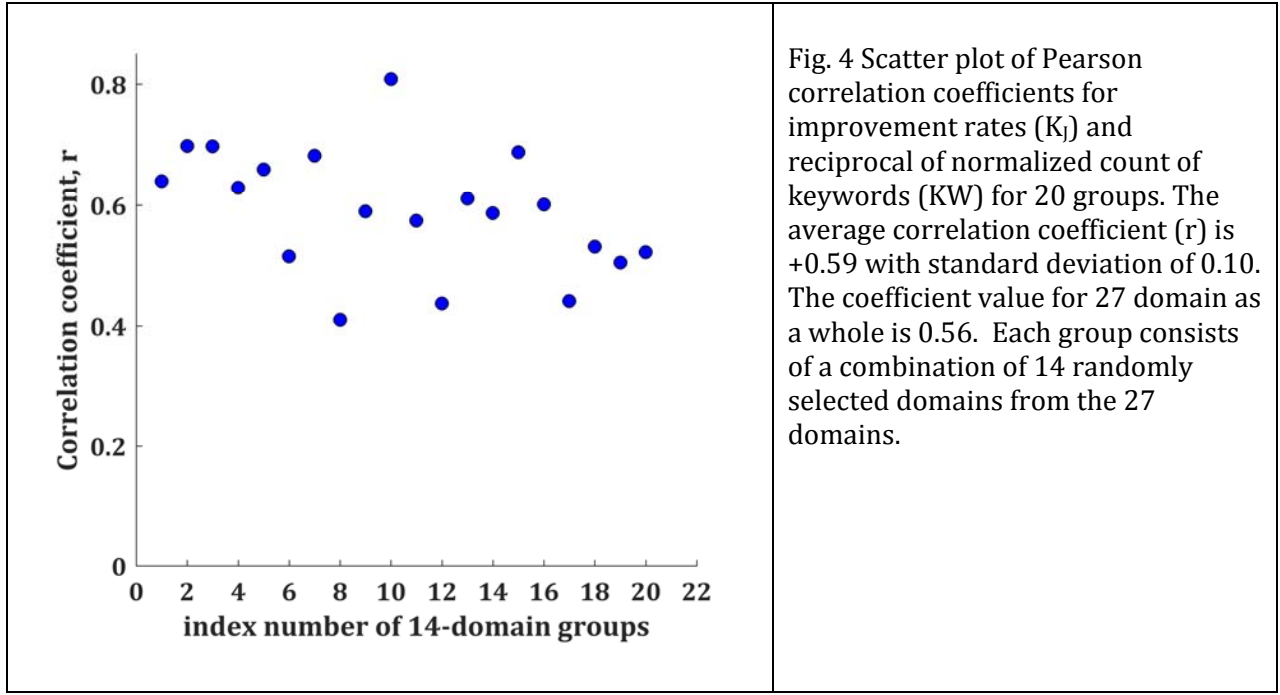

Fig. 4 Scatter plot of Pearson correlation coefficients for improvement rates ($K_J$) and reciprocal of normalized count of keywords (KW) for 20 groups. The average correlation coefficient (r) is +0.59 with standard deviation of 0.10. The coefficient value for 27 domain as a whole is 0.56. Each group consists of a combination of 14 randomly selected domains from the 27 domains.

The results of those 14 groups are presented in Fig. 4 with the index number of each group plotted along the X-axis, and Pearson's correlation coefficient along the Y axis. It is clear from the figure that the correlation values are all positive and range relatively narrowly from +0.41 to +0.81. The average correlation coefficient value is +0.59. These results show that the correlation value is relatively stable, and further supports the conclusion that the correlation coefficient that was obtained in the initial study (+0.56) was not due to random effects associated with particular domains.

## Discussion

The goal of this empirical study was to test the theoretical prediction in Equation (2) that interactions give rise to variation in performance improvement rates of technological domains, and those technological domains associated with higher levels of interactions improve at a slower rate than those with lower levels. The normalized count of selected keywords reflecting the concept of interaction as described in previous literature (11, 12, 13, 15, 16) was used to estimate artifact interactions. The correlation study found that performance improvement rates are positively related to the inverse of the normalized keyword counts.

The first finding of the empirical study is that patents can be a useful resource for studying artifact interactions, and to our knowledge, this is the first time patents have been used to study interactions. Although techniques such as design structure matrix (DSM) have given some results on interactions (17), working with DSM is challenging. Available DSM on domain artifacts is very limited; it is very expensive to develop these matrices; and, the reliability of the results is highly dependent on the researcher doing interviews and on the knowledge/memory of the people interviewed. In contrast, patents afford a large objective data set, both being publicly available, and spanning long periods (digital patents available from 1976 to the present and non-digital long before that). The selected keywords signal the different types of interactions: component-to-component, component-to-system, side effects, and coupling of functional requirements (*Material and Methods*). The keywords have been selected so that they have at least a moderate level of occurrence, and are widely used across domains hence requiring them to be non-domain specific. In a sense, keywords may be viewed as describing interactions at a higher abstraction level. For an example, usage of the keyword 'prevent' implies prevention of some problem, but without specifying what the problem is. The further specification would potentially make it domain specific and not useful in the kind of cross-domain test performed here. Additionally, the selected keywords in the text need to signal interactions most of the time (true positives), that is, they need to be relevant (*Material and Methods*). The work reported here shows that such words can be found and the six utilized here are demonstrated to be applicable to a fairly diverse range of 27 technological domains.

The analysis showed that improvement rates and inverse of normalized count keyword are positively correlated with the Pearson coefficient (r) equal to +0.56, and p-value equal to 0.002. The correlation is medium, and does not explain all the variation in the rates. This can be understood in the context of relationship (2). According to the model, another factor contributing towards variation of improvement rates is the scaling of design variables. Assuming that the influence of scaling ($A_J$) on improvement rates across the 27 domains is similar to that of the interaction parameter, Equation (2) implies that improvement rates and inverse of keywords by itself can be expected to have medium correlation.

Although this work is the first of its kind in using patents to study domain interactions, the approach has some limitations. First, it can be observed that the noise is quite significant. Although much of this could be due to the other missing theoretical variable, scaling or to inaccuracy in measuring $K_J$, another very likely source is due to limited resolution of the keywords as an estimate of interactions. An additional possible source of noise might be due to the limited number of patents (100 most-cited patents) being used for the study (due to limited resolution of Classification Overlap Method (COM) in its current state (*Material and Methods*)). This issue should be more significant for domains with low normalized count of keywords, which is consistent with the higher spread of data points at lower count of keywords in the graphs (see right side Fig. 3). An open issue is whether the keywords we have identified would work to reliably estimate interactions in domains that we have not examined. Although the keywords were carefully selected to be general, we did find that the estimate was distorted in a domain with extensive use of chemical symbols. In this study, the only problematic domain was genome sequencing but other domains could have this specific issue or other text "anomalies" that badly interfere with obtaining a meaningful estimate. Further research on more domains including reading patents would increase our understanding of the reliability of the method.

Overall, the correlation analysis from this empirical study strongly supports the theoretical suggestion that the domain interaction parameter is a factor that can lead to variation in improvement rates, where a higher interaction parameter leads to lower improvement rates. Further, it also supports the relational form the model predicts.

## Material and Methods:

**Approaches to study interactions in artifacts:** Two potentially applicable approaches exist for studying artifact interactions. One method that is well recognized is the design structure matrix (DSM) (17), which when applied to products captures interactions between components in any artifact. The empirical method utilizes interviews with a broad variety of engineers who are knowledgeable about development of an artifact and are associated with effort on various components or systems that make up the artifact. Such interviews can capture geometrical, energy, material and information interactions and the DSM can be defined at different levels of abstraction of the product and the method has been well developed for some time now. If one could obtain reliable DSM data across a wide range of domains, this would be an effective way to study interactions. However, it is very expensive to develop a DSM for a complex product such as a jet engine, an aircraft, and a MRI machine. Perhaps for this reason, the number of DSM publicly available in papers and at websites is limited. Further, it will be very hard, perhaps even impossible, to develop DSM of artifacts that were designed some time ago such as 30 years. Due to the scarcity of available data, (only one DSM is apparently available for a domain where the improvement rate is known), and prohibitive cost of developing them, this approach was not pursued further to test the influence of interactions on improvement rate.

Another approach is to use documents: One set of documents potentially relevant for studying interactions are design manuals and engineering books related to specific domains, which could be analyzed using text mining techniques. Hommes and Whitney (2003) have used documents describing product, and sub-systems level requirements to pursue case studies of system level interactions. Since very few documents describing interactions in artifacts belong to domains for which we have performance data, this was not a viable approach either.

Another set of promising documents are patents. Patents are particularly attractive; as a data source, they are generalizable, objective, and publically available. They provide meta-data, and qualitative data (drawings and text). The textual data describes state-of-art prior to the inventions, and associated problems that were solved. Second, the data is available for many generations of inventions, and easily accessible from USPTO or other websites such as Google.com. Additionally, the Classification Overlap Method (COM) (18, 19), a recently developed tool based on UPC and IPC classification codes, enables identification of patents for each specific technological domain. Unlike in DSM of products in which interactions have already been identified, no interactions, however, are inherently defined in patents as patents are written for the protection of intellectual property, and patentability does not require interactions to be identified. Thus, it was necessary to develop a method for identifying and extracting interactions.

**Procedure for text mining and analysis of patents:** The patent analysis using a text mining approach was conducted in two phases. In the pilot study, using patents from 5 domains - battery, wind power, solar PV, capacitors and computer tomography scanning (CT scan), feasibility for extracting data about artifact interactions from patent text was explored using Latent Semantic Analysis, Latent Dirichlet Analysis, and keyword-based techniques. Only the keyword-based technique was found to be useful in extracting data on interactions, and hence will be discussed further. In the extended study, the keyword-based technique was implemented in 27 domains.

Both the pilot and extended study consist of four broad steps: (1) preparation of textual data from patents (2) identification of interactions and associated keywords (3) keyword-based text mining (4) analysis and interpretation of variations in keywords across domains. In step 1, domain patents were identified, electronically retrieved from the web, and cleaned to prepare for analysis. In step 2, patents were read in detail to identify potential generic keywords associated with interactions. In step 3 the raw data on interactions was extracted using keyword-based text mining, which was then analyzed in step 4. Subsequent sections describe these four steps. Since step 2 is the most critical among these, we describe this step first.

**Identification of artifact interactions and associated keywords in patent text (step 2):** Based on the qualitative work of Simon (10), Suh (11), Weber (12), and Whitney (15, 16), we have classified interactions into three broad classes of interactions, which provide a useful framework for identifying text describing interaction and associated keywords.

**Classes of interactions:** (1) Between functional requirements: These interactions are consequences of the dependencies between multiple functions and design parameters (12). For example, increasing the size of a mechanical component can increase its stiffness, a desirable quality. But, increasing size results in increase of mass, which can affect dynamics of the artifact adversely. When one function is improved, such interactions can lead other coupled functions to be adversely affected.

(2) Between component and component, or between component and system: A good example of this type of interaction is the necessity to match impedance between sub-systems in order to maximize power transfer (15, 16).

(3) Parasitic/side effects: These represent undesirable effects exhibited by the components and sub-systems, while they fulfill their main functions (15, 16). Some examples of these are corrosion in battery electrodes, and heat dissipation in computers.

**Identification of text and keywords capturing interactions:** Using the above framework, as part of step 2 in the pilot study, two researchers, including the current lead author and an Intern working with the author for several months, read a set of 60 patents from the 5 domains noted earlier (battery, capacitor, wind power, solar PV, and CT scan) to identify text describing technical issues that reflect interactions. Three patents from each decade starting from the 1970's until the present were selected to make a total of 12 patents in each of the 5 domains. It was observed that background or prior art sections, as expected, described problems with the state-of-art artifacts. It was also found that many patents, while summarizing the current invention, also discussed problems that were not previously discussed in the background or prior art section. In both of these sections, descriptions of problems consistent with the prior research (11, 12, 13, 15, 16) were noted as interactions. The detailed description and claims sections focused on describing the current invention and novelties inventors wanted to claim as assignee's intellectual property, and rarely included descriptions of interactions. Based on this reading, the decision was made to include text from the *title, abstract, background, and summary* sections, and not include the detailed description and claims sections to maximize signal to noise ratio.

The essential part of step 2 was to extract text samples that contained the description of the interactions. These text samples were examined for keywords that tended to appear as signals of interactions but that were not domain specific. Two examples of such text (italicized) indicating interactions with associated keywords in bold are presented below in Table 1, set 1, and an extensive list is in the SI. Below each example text, our interpretation of the interaction using the framework and typology described above is presented. The first example text in the table may be interpreted as an interaction of functional requirements, specifically energy storage and preservation of electrical waveform. The second example clearly describes a side effect, where leakage of electrolyte leads to failure.

**Table 1 Examples of text from patents describing interactions, and associated keywords.**

**Set 1: Example text indicating interactions**

- *Generally, conventional aluminum electrolytic capacitors have an energy storage value or capacitance which increases with applied voltage. This is probably due to penetration of the liquid electrolyte into the aluminum oxide surface coating on the anode. Sometimes, however, such penetration is **undesirable**, as it can result in a change in the dielectric characteristics and hence in a distortion of the waveform in pulse applications.*

    Interpretation of interaction: One function of this device – a capacitor - is to store energy, and another function is to preserve the waveform (applicable to some applications). The penetration described leads to increase in voltage, which improves the energy storage capability (first function); but at the same time, it deteriorates the second function, thus causing **interaction of the two functions**. The deterioration may also be viewed as an undesirable side effect.

- *In such electrolytic capacitors there exists the risk that the liquid electrolyte will leak out. Accordingly, the capacitor must be hermetically sealed to **prevent** any leakage of the liquid electrolyte therefrom, since if the liquid were to come into contact with the other electronic components encapsulated in the device, it could damage them sufficiently to cause the device to fail to operate properly.*

    Interpretation of interaction: The electrolyte which has leaked attacks neighboring

> components causing failure. Thus this leakage and failure are an **undesirable side effects**. The keyword that reflects this side effect is the word '*prevent*'.
>
> **Set 2: Example of keyword usage *not* describing an interaction (irrelevant)**
>
> - *A primary **problem** with most CT methods is that they are time consuming. Consequently, prior to this invention, CT technology has not been a feasible alternative to such **problem**s as screening luggage for concealed items. Screening luggage for concealed items is of vital importance. Such monitoring is necessary to avoid smuggling of drugs and to detect explosives planted in luggage by terrorists. Present techniques for screening luggage include manual inspection. Manual inspection is a time consuming and therefore expensive operation.*
>
>   Explanation: This example text describes an unsatisfactory performance of the CT methods referenced by this patent, specifically slow speed of scanning. The usage of the keyword '*problem*' does not indicate an interaction. Second usage of the keyword is for an application which may be seen as a design opportunity.

From reading of the 60 patents, a total of 30 keywords were found that potentially indicated interactions. These keywords were used to study 430 additional patents from 5 domains for interactions. Using the raw data from text mining of these patents, three criteria were used to cull the keywords: (1) count of occurrence, (2) cross-domain usage of the keywords across the domains, and (3) relevancy of keywords in reflecting interaction. First, high occurrence of a keyword was necessary to get a statistically strong signal capable of showing variation across the domains. For example, the words '*problem*' and '*prevent*' were common keywords in description of technical issues. Second, since the goal of the study is to conduct a comparative study, it was also necessary to ensure that keywords were not domain specific but instead generally used. For this reason, the word '*corrosion*' was not considered a good keyword, since it would be too specific to particular domains, and may see no usage in many domains. The word '*prevent*' or '*undesirable*' was selected as a better choice in such instances, since it captures the notion of bad side effect that needs to be mitigated, but without being limited as to the detailed nature of the side effect. Third, it was also important to ensure that the keyword when it was used in text reflected interaction most of the time, that is the keyword has high sensitivity. This is assessed by relevancy of a keyword, which estimates how often a specific keyword reflects interaction when it is used in the text. This is defined as a ratio of count of keywords signaling interactions to the total count of the keyword usage in the patent set for a domain. Example patent text in Table 1, set 1 provides examples of relevant usage of the keywords. For comparison, the patent text in set 2 provides an example of non-relevant usage of the keyword 'problem'. The keyword 'problem' in this example is used to indicate inadequate performance and design opportunity, not interactions.

Data from the pilot study of 430 patents showed that majority of the 30 keywords words had low count, and hence were removed from the list. Additionally, using the cross-domain usage criteria, the remaining keywords were reduced to 8 keywords: ***parasitic***, ***problem***, ***prevent***, ***undesir***able, ***requirement***, ***fail***ure, ***disadvantag***e, and ***overcom***e. The root of each keyword searched is shown as bold and italicized text. The root is chosen so that the searched keyword is able to detect maximum number of instances of the keyword, but without reducing the relevancy percentage. These 8 keywords had high count and are relatively general and not domain specific.

Relevancy percentages for the keywords were determined from reading of the 60 patents (described above) for identifying interactions and associated keywords. Mean relevancy percentages for these 8 keywords across the 5 domains are listed here inside the parentheses: *parasitic (0.97), problem (0.58), prevent (0.83), undesir*able (0.94)*, requirement (0.75), fail*ure (0.72)*, disadvantag*e (0.81)*,* and *overcom*e (0.98). The relevancy percentage for each individual domain is presented in Table S3 in SI. The keyword 'overcome' had indicated interactions in nearly all instances (on average 98 % of the times). In contrast, the keyword 'problem' has poor relevancy percentage; almost half the times (42%) it did not indicate interactions. This is because it is common among engineers to use the word '*problem*' to describe design opportunities to improve main functions of a technology, or to take advantage of new applications. This convention is reflected in the patent text as exemplified by the example in Table 1, set 2. Because of the low relevancy percentages and its possibility to add much noise to the data, the keyword '*problem*' was removed from the list.

The remaining 7 keywords were further vetted in the extended study with patents from 27 domains. The raw data from keywords showed that the keyword 'parasitic' did not have wide cross-domain usage. In fact, 12 domains did not use it even once, and only 5 domains – Camera Sensitivity, Capacitor, Electric Power Transmission, Fuel Cell, IC chips - used it often. Fig. 5a shows the distribution of the "parasitic" keyword across all patents in 27 domains. To provide perspective, the distribution of the keyword "prevent" is presented in Fig. 5b. It is clear the keyword "prevent" is used by all domains frequently. Due to a low cross-domain usage, the keyword 'parasitic' was also eliminated, leaving with a final list of 6 keywords for cross-domain analysis.

**Preparation of text from domain patents (step 1):** The outcome of this step was the raw text from the 100 most-cited patents for each of the domains being studied. The following procedure was used to retrieve the text: (1) Patents in each specific domains were identified using the COM method (19, 20); out of which one hundred and fifty most-cited patents[‡] in each domain were chosen and read to eliminate any irrelevant patents from the set; then the 100 most-cited relevant patents were selected from the patent set. The patents had been identified as part of doctoral research (21) in which both authors of this paper participated in reading the patents for relevancy. (2) Four sections of the text – title, abstract, background, and summary – in each patent were downloaded from Google's patent database. (3) To prepare the text for mining, extraneous text -such as stop words- was removed using Python scripts. Stop words are a set of commonly used words, such as *the, a, it,* and *in*. Although they are critical in natural language, they do not add any value to the data. Removing them makes it possible to focus on the important words, and to reduce computational cost as well as noise in the data. The completion of this step prepared the text for mining.

---

[‡] The most cited 150 patents were obtained so that 100 could be retained after eliminating non-relevant patens. This relevancy is different from the relevancy of keywords for indicating interactions.

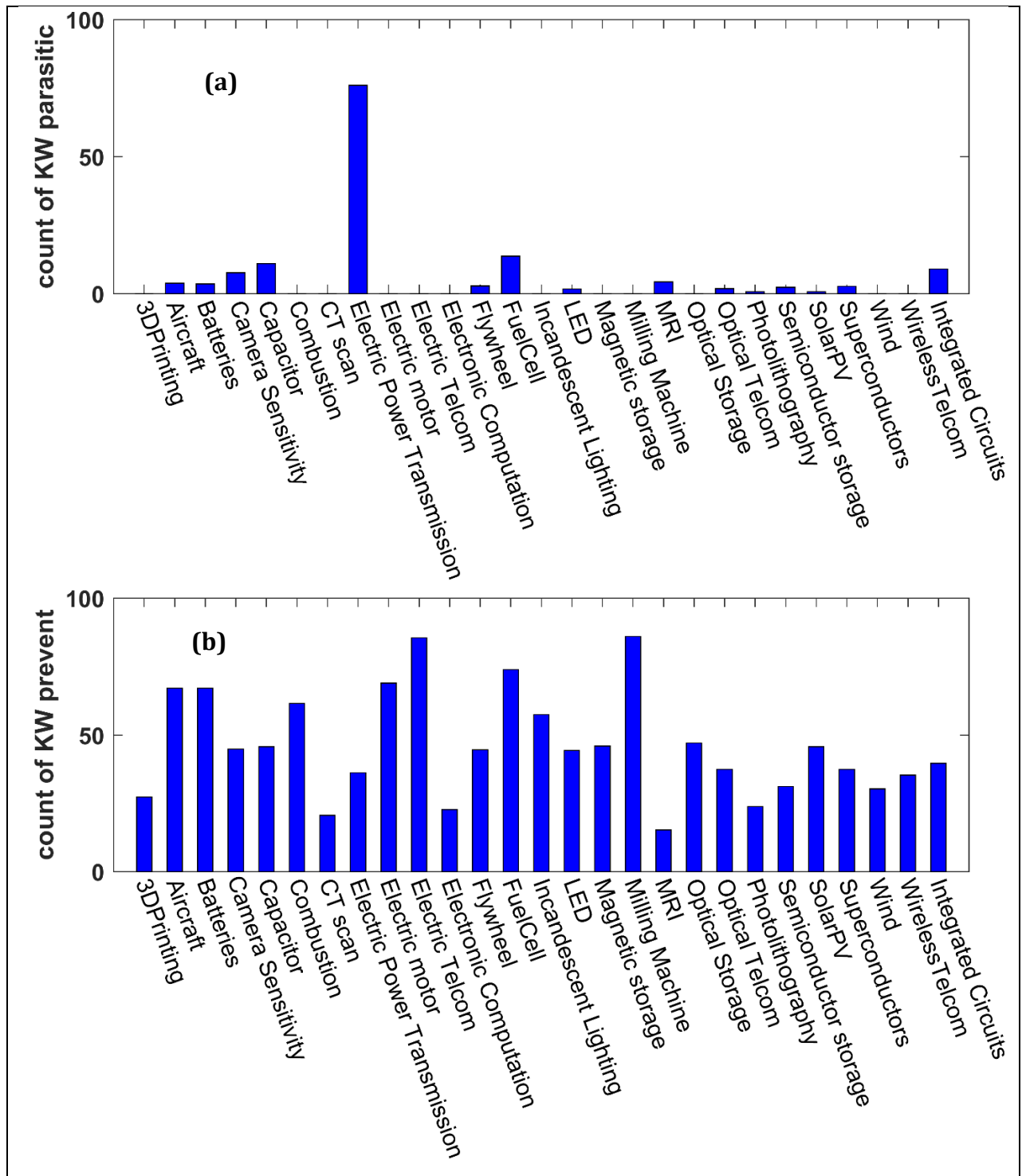

Fig. 5 Cross-domain usage of keywords (KW) 'parasitic' (panel a) and 'prevent' (panel b) across 27 domains. Keyword 'parasitic' is not widely used; in fact, 12 domains do not use the 'parasitic' keyword at all. Compare this to wide cross-domain usage of keyword 'prevent'. The count of KW presented is a normalized count of the respective keyword, with normalization against 100 thousand total words.

Out of the 2700 patents used for extended study (sub-step 2 above), 2400 patents were downloaded using a web-scraping tool from Google patent database, which provides patents in searchable html files. The remaining patents had to be downloaded manually.[§] Almost all of these manually downloaded patents either lacked proper titles describing the sections, or background and summary were merged with detailed descriptions. For these cases, the background information and summary had to be manually identified by reading the patents and extracted.

**Text mining (step 3) and analysis (step 4):** The prepared patent text for the 27 domains was mined using Python scripts for determining the count of the keywords as well as the total number of words in the patent set in each domain. The total number of words in the patent set was used for normalizing the count of keywords in each domain, and the normalized keyword count is expressed per 100 thousand words in the patent set.

For analysis, we assumed a linear relationship between normalized count of keywords and artifact interactions described by the models (8, 9). We then tested dependence of improvement rates on artifact interactions as predicted by the models by conducting a correlation analysis of improvement rates and normalized keyword count from the patent text for the 27 domains.

## Acknowledgements:

The authors are grateful to the International Design Center of MIT and the Singapore University of Technology and Design (SUTD) for its generous support of this research. We want to also acknowledge valuable input on an earlier version of this paper by Dr. James McNerney and Dr. Daniel. E. Whitney.

---

[§] Since 24 patents could not be downloaded, the number of patents in each domain study ranged from 97 to 100.

## Supplemental Information (SI)

The Supplemental information provides examples of patent text with keywords used to estimate domain interactions. It also presents additional information for constructing and analyzing data, as well as raw data for count of keywords and words in the patent sets.

Table S1 lists the words used for searching sections in patents available in Google's patent database available on the internet.

Table S2 provides raw data from the extended study of interactions, namely the number of patents in each domain patent set, keyword count in each domain, word count, 6 keyword count, total number of words in each domain patent set, and normalized 6 keywords per 100,000 words.

Table S3 provides data on relevancy for 8 keywords.

The largest section in the SI presents 42 examples (with 7 for each keyword) of text describing interactions with the keyword in bold. Each example also includes a brief explanation and classification of the type of interaction.

| | **Section name** | **terms used for searching section headers** |
|---|---|---|
| 1 | Title | 'Patent-title' |
| | | 'Invention-title' |
| 2 | Abstract | 'Abstract' |
| 3 | Background (for exact match in heading) | 'description of the prior art', |
| | | 'background of the invention', |
| | | 'background', |
| | | 'background information', |
| | | 'prior art', |
| | | 'introduction to the invention' |
| | Background (for partial match in heading) | '.*background.*', |
| | | '.*prior art.*', |
| | | '.*related technology.*', |
| | | '.*related art.*' |
| | Background (for partial match in paragraph) | '.*background.*', |
| | | '.*prior art.*', |
| | | '.*related art.*' |
| 4 | Summary (for exact match in heading) | 'summary of the invention', |
| | | 'statement of the invention', |
| | | 'general description of the invention', |
| | | 'brief description of the invention', |
| | | 'short description of the invention', |
| | | 'brief description of the present invention' |
| | Summary (for partial match in heading) | '.*summary.*' |
| | (for partial match in paragraph) | '.*summary.*', |
| | | '.*statement of the invention.*', |
| | | '.*general description of the invention.*', |
| | | '.*brief description of the invention.*' |

Table S1: Search terms (between quotes) used for identifying the sections in the Google patent database. Asterisk (*) represents a wild card character used in Python code for searching multiple characters in text strings.

Table S2: Summary of data from empirical study of interactions in 28 domains. Columns 1 and 2 identify the domains; column 3 lists number of patents used in domains; columns 4-9 list count of 6 keywords (*prevent, undesirable, requirement, fail, disadvantage, overcome*) followed by cumulative count of 6 keyword for each domain; columns 11 lists words in each domains, then followed by normalized count of 6 keywords, and performance improvement rate ($K_J$).

| | | | Count of each keyword | | | | | | | | | |
|---|---|---|---|---|---|---|---|---|---|---|---|---|
| Domain # | Domain name | Number of patents | Prevent | Undesirable | Requirement | Fail | Disadvantage | Overcome | 6KW, total | words, total | (6KW/ Words) * 1e5 | K% |
| Domain_1 | 3DPrinting | 100 | 47 | 14 | 31 | 11 | 45 | 14 | 162 | 172952 | 94 | 38 |
| Domain_2 | Aircraft | 100 | 88 | 14 | 81 | 99 | 48 | 24 | 354 | 131060 | 270 | 12 |
| Domain_3 | Batteries | 100 | 75 | 8 | 48 | 28 | 18 | 15 | 192 | 111825 | 172 | 7 |
| Domain_4 | Camera Sensitivity | 99 | 58 | 3 | 34 | 3 | 25 | 19 | 142 | 129106 | 110 | 16 |
| Domain_5 | Capacitor | 100 | 54 | 23 | 30 | 25 | 32 | 16 | 180 | 117888 | 153 | 15 |
| Domain_6 | Combustion | 99 | 69 | 12 | 23 | 41 | 22 | 22 | 189 | 112038 | 169 | 6 |
| Domain_7 | CT scan | 100 | 31 | 16 | 21 | 14 | 32 | 23 | 137 | 151289 | 91 | 37 |
| Domain_8 | Electric Power Transmission | 100 | 42 | 15 | 48 | 28 | 32 | 13 | 178 | 115704 | 154 | 15 |
| Domain_9 | Electric motor | 99 | 66 | 20 | 29 | 13 | 42 | 21 | 191 | 95661 | 200 | 3 |
| Domain_10 | Electric Telcom | 100 | 88 | 9 | 36 | 25 | 26 | 16 | 200 | 102817 | 195 | 10 |
| Domain_11 | Electronic Computation | 99 | 33 | 0 | 58 | 62 | 19 | 9 | 181 | 146260 | 124 | 33 |
| Domain_12 | Flywheel | 100 | 48 | 11 | 39 | 80 | 54 | 23 | 255 | 107438 | 237 | 9 |
| Domain_13 | FuelCell | 99 | 108 | 14 | 73 | 11 | 28 | 17 | 251 | 146123 | 172 | 14 |
| Domain_14 | Genome sequencing | 99 | 42 | 2 | 16 | 7 | 21 | 13 | 101 | 191484 | 53 | 29 |
| Domain_15 | Incandescent Lighting | 100 | 63 | 15 | 21 | 62 | 42 | 14 | 217 | 109610 | 198 | 5 |
| Domain_16 | LED | 100 | 53 | 7 | 12 | 16 | 29 | 17 | 134 | 119257 | 112 | 36 |
| Domain_17 | Magnetic storage | 99 | 64 | 7 | 43 | 17 | 26 | 29 | 186 | 139223 | 134 | 32 |
| Domain_18 | Milling Machine | 97 | 89 | 16 | 28 | 22 | 37 | 28 | 220 | 103482 | 213 | 3 |
| Domain_19 | MRI | 98 | 21 | 14 | 24 | 17 | 58 | 20 | 154 | 138033 | 112 | 48 |
| Domain_20 | Optical Storage | 99 | 72 | 3 | 19 | 34 | 31 | 9 | 168 | 152731 | 110 | 27 |
| Domain_21 | Optical Telcom | 99 | 40 | 7 | 31 | 6 | 23 | 22 | 129 | 106801 | 121 | 65 |
| Domain_22 | Photolithography | 98 | 33 | 27 | 31 | 11 | 13 | 14 | 129 | 139494 | 92 | 24 |
| Domain_23 | Semiconductor storage | 97 | 41 | 7 | 28 | 30 | 47 | 21 | 174 | 132235 | 132 | 43 |
| Domain_24 | SolarPV | 98 | 59 | 11 | 42 | 25 | 22 | 13 | 172 | 128842 | 133 | 10 |
| Domain_25 | Superconductors | 100 | 41 | 14 | 20 | 11 | 15 | 19 | 120 | 109385 | 110 | 10 |
| Domain_26 | Wind | 99 | 39 | 8 | 31 | 29 | 34 | 29 | 170 | 129593 | 131 | 9 |
| Domain_27 | WirelessTelcom | 99 | 52 | 8 | 60 | 19 | 33 | 29 | 201 | 147087 | 137 | 50 |
| Domain_28 | Integrated Circuits (IC) | 99 | 44 | 11 | 54 | 13 | 37 | 14 | 173 | 110844 | 156 | 36 |

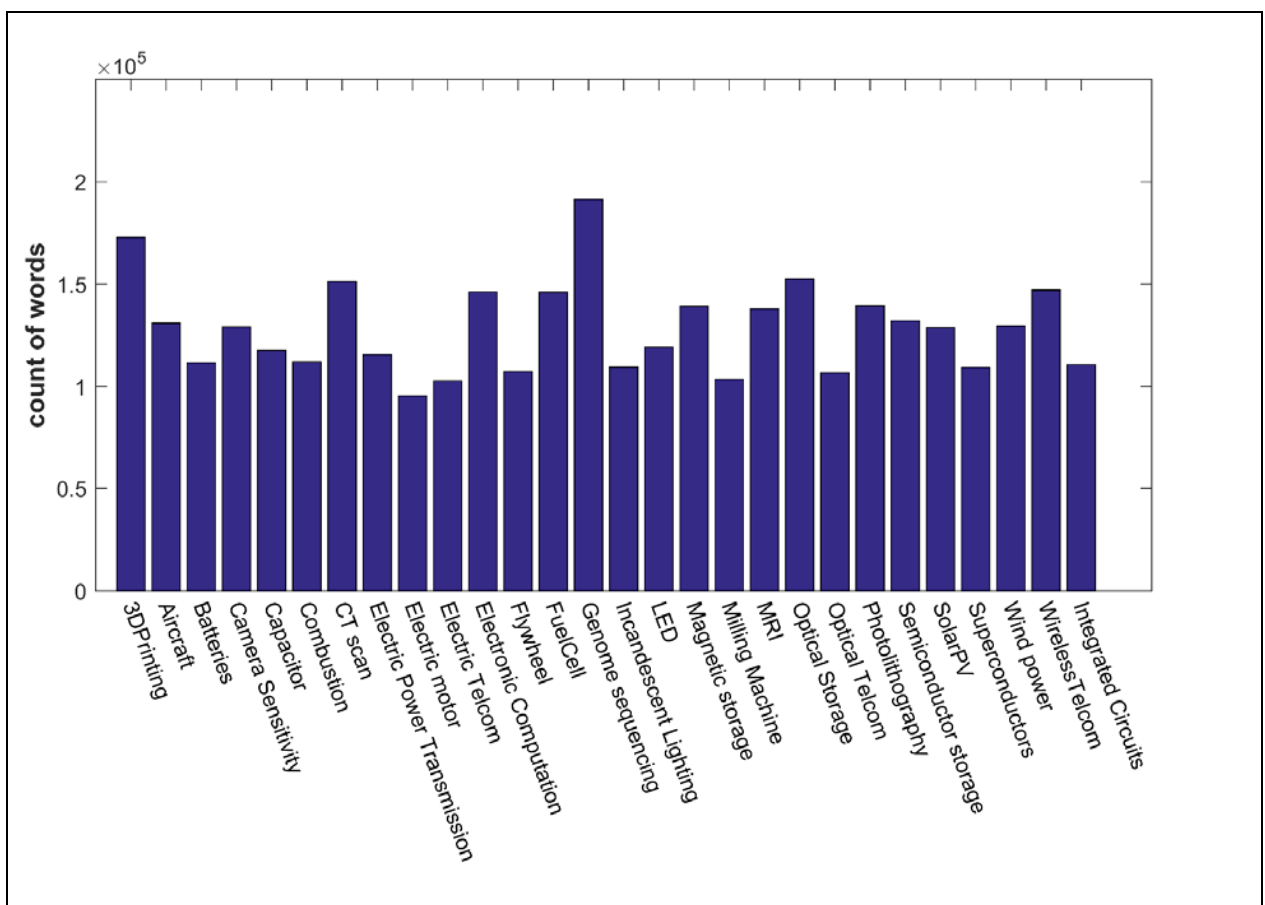

Fig. S1 Variation of count of all words per domain. The count of words includes count keywords and all other words in the text (abstract, title, background, and summary of invention) from approximately 100 most-cited patents in each domain. This does not, however, include the count of stopwords (e.g., articles etc.) removed from the text.

Table S3: Relevancy of keywords in 5 domains, and their average across 5 domains; / indicates no keywords were found in the text studied

| Keywords | Arithmetic mean | Batteries | Wind | PV | Capacitors | CT scan |
|---|---|---|---|---|---|---|
| *parasitic* | 0.97 | 0.99 | / | 0.99 | 0.94 | / |
| *problem* | 0.58 | 0.68 | 0.61 | 0.55 | 0.5 | 0.55 |
| *prevent* | 0.83 | 0.93 | 0.85 | 0.76 | 0.85 | 0.75 |
| *undesir*able | 0.94 | 0.88 | 0.95 | 0.99 | 0.99 | 0.875 |
| *requirement* | 0.75 | 0.85 | 0.69 | 0.79 | 0.81 | 0.63 |
| *fail*ure | 0.72 | 0.92 | 0.68 | 0.78 | 0.74 | 0.5 |
| *disadvantag*e | 0.81 | 0.93 | 0.8 | 0.79 | 0.8 | 0.71 |
| *overcom*e | 0.98 | 0.99 | 0.96 | 0.99 | 0.95 | 0.99 |

Relevancy percentages shown in Table S3 for the keywords were determined by independent reading of the 60 patents by two researchers for identifying interactions and associated keywords. The mean relevancy percentage for each keyword in a specific domain was calculated as a ratio between count of the keyword indicating interactions (true positives) to the total count of the keyword in the patent set of the specific domain. The mean relevancy percentage for each keyword is the average value of relevancy percentages from the 5 domains.

**Additional examples of patent text using the six keywords**

For each of the final 6 keywords (*prevent, undesirable, requirement, fail, disadvantage, and overcome)*, 42 additional example text are presented below for illustration purpose only. These examples were selected in the following manner: a domain was chosen randomly, then one patent out of the possible 100 was randomly chosen and then example text associated with the keyword was randomly chosen within the corpus of the domain starting at the chosen patent. In a few cases, where the examples required considerable analysis and excessive text length to see whether the text described an interaction, text from the next occurrence is presented to make it easier for the reader. Thus, the selected examples show slightly lower occurrence of non-relevant text (no interaction) than indicated in Table S3.

Forty-two examples, seven for each key word, are presented. These examples are in addition to the three shown in the Methods and Materials section. Each example text starts with the technological domain name, the patent number, and the title of the patent from which the text was retrieved; the example text which includes the keyword in bold text follows immediately. The type of interaction (between functional requirements, between component and component, or between component and system, or parasitic/side effects) is identified for each example. It should be noted that some of the interactions can be interpreted in two alternative ways by this analysis but we present only the interpretation clearest to us. The reading of the following examples, while perhaps somewhat subjective and possibly not statistically significant, indicate that out of the three interactions, side effect has the highest frequency (22 out of 42), functional requirement comes next (15 out 42), and the component-to-component and component-to-system have only a few (2 of 42). Three examples with keywords are non-relevant, and represent null cases.

**Keyword: Prevent**

1. |Battery|US5314765|Protective lithium ion conducting ceramic coating for lithium metal anodes and associate method|

*Such cells relied, for the most part, on separator structures or membranes which physically contained a measure of fluid electrolyte, usually in the form of a solution of a lithium compound, as well as providing a means for **prevent**ing destructive contact between the electrodes of the cell.*

Interpretation of interaction: destructive contact between the electrodes (two components) is a **harmful side effect** that is being prevented.

2. |Integrated circuits|US7064346|Transistor and semiconductor device|

*A thin film transistor using amorphous silicon, polycrystalline silicon or the like has been generally used as a transistor for use in driving liquid crystal display devices. Since these materials exhibit photosensitivity for the visible light region, carriers are generated by a beam of light, and resistivity of a thin film constituting the thin film transistor is lowered. For this reason, when the beam of light is radiated thereonto, the transistor may be made to be a turn-on state, in spite of the fact that the transistor must be controlled to be a turn-off state. Accordingly, to keep the transistor at the turn-on state, the lowering of the carrier resistivity of the thin film due to the radiation of the beam of light has been heretofore **prevent**ed by the use of a light shielding layer made of a metal film or the like.|*

Interpretation of interaction: The transitions of the npn material control the resistivity of transistor. However, this resistivity is compromised by a parasitic effect that exhibited due to photosensitivity of npn material. This parasitic effect is the **harmful side effect**.

3. |Electric motor|US4916340|Movement guiding mechanism|

*As an important feature, the moving element of the linear motor or the stator of the motor is coupled to an associated component by use of a specific spacer, and the interspacing defined between the moving element or the stator and its associated component is filled with a heat insulating material. With this arrangement, the heat transmission from the linear motor to the surface plate is effectively **prevent**ed without interfering with the transmission of a drive force of the linear motor.*

Interpretation of interaction: The text implies there is both transfer of drive force, and transfer of heat, when there is no spacer. The heat insulating spacer enables passage of drive force minimizing transfer of heat, which is a **harmful side effect**.

4. |Fuel cell|US4642273|Preventing irregular heating|

*When a load of a fuel cell to which hydrogen gas is supplied from a fuel reformer reactor is suddenly decreased, the temperature in the reformer reactor is rapidly increased, thereby damaging the construction materials thereof. In order to solve this problem, a controlling system for the reforming reaction transitionally reduces the pressure in the reformer reactor when the sudden decrease of load takes place, so as to increase the reaction rate in said reformer reactor, and to increasing the quantity of endothermic heat by said reaction, thereby controlling the rapid increase of temperature in the reformer reactor.| This invention relates to a control system for a reformer reactor which supplies hydrogen to a fuel cell. Particularly, this invention relates to a control system to **prevent** irregular heating in a reformer reactor.|*

Interpretation of interaction: When the load is reduced the temperature increases rapidly, in the reformer. The resulting high temperature damages the construction materials of the reformer. The construction material of the reactor exposed to this heat is damaged. The interaction between gases in the reactors and construction materials can be viewed as **component-to-component interactions**.

5. |Camera Sensitivity|US4952788|Method of photoelectric detection with reduction of remanence of a phototransistor, notably of the NIPIN type|

*The disclosure concerns photosensitive matrices, and especially those using NIPIN or PINIP type phototransistors made of amorphous silicon. To **prevent** problems of remanence, due to the collecting of holes in the base after an illumination stage, it is proposed to follow the step for reading the illumination signal by a remanence erasure step in which the phototransistor is made conductive in forward or reverse bias, so as to inject, into the base, electrons which will eliminate the holes by recombination. Switching on by reverse bias proves to be more efficient than switching on by forward bias.*

Interpretation of interaction: Remanence is an **undesirable side effect** experienced by the device when illuminated; the erasure step is a countermeasure to eliminate the effect of the remanence.

6. |Magnetic Storage|US5034837|Magnetic disk drive incorporating a magnetic actuator lock and a very small form factor|

**Prevent***ion of this `crash` or `stiction` phenomenon depends largely on strict control of the disk surface quality and of the chemistry, consistency and deposition of surface lubricant. This is difficult and expensive process for the typical disk drive…*

Interpretation of interaction: Stiction is an **undesirable side effect** experienced between the desirably close reading head and rotating disks in disk drives.

7. |3DPrinting|US5545367|Sequential polymerization of polymer precursor fluids; minimum stress between layers and low curl distortion|

*The still-liquid material ahead of the moving polymer zone can then flow freely, at a rate that equals the rate of shrinkage, and a distortion-free, reduced stress polymeric network is produced. Using this process, objects can be cast in a way to **prevent** cavitation, or voids caused by the shrinkage of material during polymerization. This method is referred to below as the ""sequential polymerization"" modification of the three dimensional stereolithographic process.*

Interpretation of interaction: Cavitation and voids are flaws that can be interpreted as **harmful side effects** of the desired polymerization process.

**Keyword: Undesirable**

8. |Batteries|US6300002|Notched electrode and method of making same|

*One solution to this problem has been to introduce an insulator ring around the outer circumference of the top of the electrode assembly to prevent contact between the extended edge of the positive electrode and the groove. However, the insulator ring tends to push the positive electrode into contact with the sidewall of the can, producing another unintentional and **undesir**able opportunity for internal shorting. There is therefore a need for an electrode that avoids unintended contact with the cell container.|*

Interpretation of interaction: The desired result in this device is to have no contact between the positive electrode and the groove (first function) and no contact between the positive electrode and the side wall (second function). The solution, placement of a ring, improves the first function but deteriorates the second function. Thus, this interaction can be interpreted as interaction between functions.

9. |Camera Sensitivity|US4853785|Electronic camera including electronic signal storage cartridge|

*CCDs are highly photosensitive and are capable of providing high resolution images. However, CCDs are relatively small in size; the typical CCD array being a two dimensional matrix approximately one centimeter square. The largest CCDs currently produced are one dimensional arrays no greater than approximately 3 to 4 inches in length. Obviously, these size constraints impose restrictions on the utility of CCDs for electronic photographic applications. It is to be noted that, while CCDs generally provide a high degree of resolution, in order to have commercial impact and to be of practical utility for photographic applications, an optical reduction system must be employed. Since such optical systems project a reduced*

*size image of the object being photographed onto the surface of the CCD, said optical reduction systems have the undesirable affect of effectively decreasing the resolution of the CCD. The optical system, itself, degrades image resolution to some degree, but, the actual reduction process is the factor which most severely degrades the effective resolution of the image formed by a CCD. Since such optical systems project a reduced size image of the object being photographed onto the surface of the CCD, said optical reduction systems have the **undesir**able affect of effectively decreasing the resolution of the CCD. The optical system, itself, degrades image resolution to some degree, but, the actual reduction process is the factor which most severely degrades the effective resolution of the image formed by a CCD.*

Interpretation of interaction: The optical system reduces the size of the image to fit the size of CCD (first function), so that it provide commercial utility; but this solution reduces the resolution of the images (another desirable functional output) in the process of reducing the size of the image. Thus, there is **interaction between two functions** – size reduction and resolution of images.

10. |Electric motor|US5208497|Linear driving apparatus|

*When an exciting current is supplied to the movable member 71 and a steering force F is exerted, a reaction force Fr (Fr=-F) works on the magnetic field forming member 73 as a counter-effect of the steering force F, which is transmitted to the base frame 74, thereby causing **undesir**able results such as vibration or deformation of the base frame 74. The above defect provides a particularly significant problem in the case where the linear driving apparatus is applied to the copying machine or magnetic disk apparatus.*

Interpretation of interaction: The vibration and deformation is an **undesirable side effect**.

11. |Flywheel|US6138527|Methods for making a flywheel|

*The flywheels proposed by both Place and Poubeau satisfy the need for high inertial volumetric packing; that is, they place the majority of the flywheel weight at the outer*

*periphery. The flywheels are also able to radially decouple from the hub assembly. However, these designs produce high point stress loads on the wound rim and produce an **undesir**able concentration of bending strains which unduly limit the total energy storage by failure at the spoke locations. They are also subject to high dynamic imbalance problems due to the wide, variable stretch which can lead to non-uniform differential radial strains between spokes.*

Interpretation of interaction: The better inertial volumetric packing – placement of the majority of the flywheel at the periphery – increases the energy density (first function). While doing so, the design in question develops stress concentrations on the rim, which lead to failure (or reducing the durability of flywheels, a second functional output). The **interactions is between two functions** – energy density and durability.

12. |LED|US6057647|Light emitting device used for display device|

*In the process of manufacturing the organic EL element, if there is a particle (small piece of trash, dust, **undesir**ed material or the like, these will be called particle hereinafter) present, in particular, on the electrode (anode electrode) immediately before the formation of the organic EL layer, the particle remains in the interface between the anode and the organic EL layer formed thereon. In case where the organic EL layer is formed while a particle remains as described, the following problems easily occur. First, the thickness of the organic EL layer is usually 500 nm or less; however the intensity of the electric field applied in order to inject carriers to the organic EL layer becomes extremely high as about several thousand V/m. Consequently, with the presence of a particle which usually exceeds 1 m in size, short-circuiting easily occurs between the electrodes. Due to the short-circuit between the electrodes, the sections of the cathode and organic EL layer, where the particle was present, were damaged, thus causing a pin hole.*

Interpretation of interaction: The formation of pinholes in the EL layer is an **undesirable side effect** caused by the short-circuit precipitated by the particulate contaminants in the thin EL layer.

13. |Optical Storage|US5748598|Apparatus and methods for reading multilayer storage media using short coherence length sources|

*To increase optical storage capacity, storage media with multiple data storage layers and systems for reading such media have been proposed. The density of storage media having multiple data storage layers is, however, limited by cross-talk. Cross-talk occurs because optical signals reflected from the media contain both desirable reflections from the layer of interest and **undesir**able reflections from the adjacent layers.*

Interpretation of interaction: Harmful reflections from adjacent media introduces cross-talk is an example of a **component/system interaction**.

14. |Integrated Circuits|US5869843|Memory array having a multi-state element and method for forming such array or cells thereof|

*In part, this is due to the fact that these processes are performed at lower temperatures; typically below 350 C. Additionally, chalcogenide materials are very easily contaminated. Direct contact between silicon and chalcogenide can ""poison"" both materials, rendering each inoperable. For this reason, isolation barriers, or diffusion barriers, must often be included to prevent **undesir**able contamination of the chalcogenide material.*

Interpretation of interaction: Poisoning of silicon and chalconide is a **harmful side effect**, and isolation barriers are a counter measure to eliminate these side effects.

**Keyword: Requirement**

15. 0|SolarPV|US4465575|Method for forming photovoltaic cells employing multinary semiconductor films|

*This method is advantageous because it eliminates the **requirement** for substrate heat within the deposition environment. If the substrate must be exposed to heat at some point, it is far easier and less expensive to do so in a furnace, after deposition, than to heat the substrate in a* *vacuum coating chamber. Temperature-related problems of mechanical motion systems (conveyors, etc.) are particularly acute in a vacuum environment.*

Interpretation of interaction: Interaction between two functional requirements: the need for deposition and the need for heating the substrate. The patent implies that state-of-art requires both of these functions to be achieved together in the deposition chamber, and couples these functions. In other words, the **two functions are interacting**.

16. |Integrated Circuits|US5250843|Multichip integrated circuit modules|

*The process … allows a variety of materials to be used including thermoplastics and thermal sets while still maintaining a high degree of planarity in the final module; allows the use of completely flat substrates, without the **requirement** for wells or substrate frames; and allows high volume production.|*

Interpretation of interaction: There is **no clear signal for interaction** in the text which uses the keyword *requirement*. This is an example of noise introduced by a non-relevant sample.

17. |Battery|US5169736|Electrochemical secondary element|

An important impetus for this invention was the recent, great increase in demand for batteries having high energy density and low weight, such as had already been achieved particularly with the lithium systems, but which are also rechargeable. This requires, among other things, that the electrodes be chemically stable in contact with the electrolyte. Lithium electrodes do not meet these *requirement*s when in use over extended periods of time, even in organic electrolytes with an aprotic solvent, because their cycling stability is well known to be sharply limited.

Interpretation of interaction: This is another example of noise as this excerpt **does not show evidence of an interaction** despite discussing two functional outputs – high energy density and the ability to recharge.

18. |CT scan|US4599740|Radiographic examination system|

*The **requirement** for increasing mechanization in the transportation of goods has led in recent years to the introduction of standardized container units which can be moved individually by road or on the trailers of tandem trucks, and which can be stacked regularly within the holds of ships to maximize the utilization of space. Such containers are loaded and closed at their departure point and provide the not inconsiderable advantage of being relatively inviolate such that loss of contents due to misplacement of the goods (or pilfering) is substantially reduced. Although the enclosure of goods within such containers to make them inaccessible is an advantage in some respects, it also means that supervision of the movement of goods for customs and excise purposes, and for verification of the contents by the appropriate Authorities at ports and land border posts is made more difficult. This is particularly troublesome if the containers are packed with a large quantity of rather small items, because the unpacking of such a container (and these can be up to 40 feet long and 8 feet width and height) creates substantial logistical problems. Problems of investigating the interior of otherwise closed containers or items which for one reason or another cannot be opened have arisen in other situations.*

Interpretation of interaction: The shipping containers provide easy and fast movement of goods both in ships and trucks (one function) and minimize pilfering. On the other hand, they make access to goods (second function) for inspection at the customs and ports difficult. This is an **interaction of functional requirements**.

19. |CT scan|US6542580|Relocatable X-ray imaging system and method for inspecting vehicles and containers|

*… relocatable inspection systems have been developed that can be assembled and used at a variety of locations to inspect large commercial vehicles and cargo containers. These systems, however, currently have many shortcomings. Specifically, existing relocatable X-ray inspection systems are extremely cumbersome to transport from one location to the next, and*

*they generally require lengthy disassembling and assembling procedures. Furthermore, these systems generally require powerful machinery to load and unload their components onto and off of multiple transport trucks for relocation. Thus, significant time and expense are required to transport and assemble existing relocatable X-ray imaging systems. As a result, for a given site or event requiring such an inspection system, substantial notice must be given, and substantial money expended, to allow for the time and preparation required to transport and assemble the system. This, in turn, presents significant logistical problems where an event requiring security inspections occurs on short notice. Additionally, existing relocatable X-ray inspection systems are generally designed for inspecting large trucks and cargo containers, not for inspecting passenger vehicles. As stated, the current systems are extremely cumbersome and time-consuming to relocate, and as such, are much larger than that which is required to inspect smaller passenger vehicles. Moreover, many of these existing systems have large conveyor platforms, which a vehicle driver may not readily step onto and off of without the aid of steps and/or railings. In light of the above, a need exists for an X-ray imaging system used to inspect passenger vehicles that is readily relocatable, and flexible in terms of on-the-spot reconfiguration, such that a wide variety of site **requirement**s can be met in a short amount of time and at minimal expense.|*

Interpretation of interaction: Portability is a challenge in the existing inspection systems; portability (new function) of inspection systems as well as suitability for passenger vehicles (flexibility) addressed by the new inspection system are not currently adequate. The **interaction between functional requirements** (speed and low cost of relocating against portability) has not been sufficiently resolved in past attempts.

20. |3DPrinting|US6841116|Selective deposition modeling with curable phase change materials|

*Previously expedients in SDM have attempted to dispense the materials at the highest temperature possible where the viscosity would be low enough to meet the specifications of the ink jet print heads. However, this approach does not work for UV curable materials because the high dispensing temperatures can initiate the cure process, which can increase*

*the viscosity of the material and undesirably effect dispensing. Adding low molecular weight monomers to lower the viscosity of the formulation can help; however, odor problems can arise, as these monomers tend to evaporate and condense within the machine causing contamination that can cause the machine to malfunction. Thus, including low molecular weight monomers is desirably minimized. The cross-linking problem is further complicated since the trend in the ink jet printing industry is to achieve higher printing resolution by decreasing the size of the orifices in the print head. As orifice size decrease, the viscosity* ***requirement****s for the material being dispensed must decrease.*

Interpretation of interaction: Easier flow of material in the 3D printing, a consequence of lower viscosity of the material, is a desirable functional requirement. The increasing temperature to reduce viscosity, however, can initiate curing process before the material is dispensed (proper timing of curing process, another functional requirement). Thus these **two functional requirements interact**, with flowability improving but with the curing process accelerated beforehand. One other design solution - monomer addition - also shows the interaction between the flow and another functional requirement - machine durability.

21. |Electronic Computation|US5136697|System for reducing delay for execution subsequent to correctly predicted branch instruction using fetch information stored with each block of instructions in cache|

*The time taken by a computing system to perform a particular application is determined by three basic factors, namely, the processor cycle time, the number of processor instructions required to perform the application, and the average number of processor cycles required to execute an instruction. Overall system performance can be improved by reducing one or more of these factors. For example, the average number of cycles required to perform an application can be significantly reduced by employing a multi-processor architecture, i.e., providing more than one processor to execute separate instructions concurrently. There are disadvantages, however, associated with the implementation of a multi-processor architecture. In order to be effective, multi-processing requires an application that can be*

*easily segmented into independent tasks to be performed concurrently by the different processors. The **requirement** for a readily segmented task limits the effective applicability of multi-processing. Further, the increase in processing performance attained via multi-processing in many circumstances may not offset the additional expense incurred by requiring multiple processors.*

Interpretation of interaction: Multi-processing reduces time for computation (one function) by decreasing the average number of cycles required for computation. Multi-processing can be done only if the application can be segmented; this requirement reduces flexibility of the applications (another function) to which multi-processing can be applied; and multi-processing can increase cost. These **two functions** – reduction of speed and flexibility of application being processed – **are here shown to be conflicting interactions**.

### Keyword: Fail

22. |Electric Telcom|US4998885|Elastomeric area array interposer|

*The number of electronic circuits that can be manufactured per unit area of silicon or board space has increased dramatically in recent years. This increase in circuit density has produced a corresponding increase in the number of connections required between the various electronic circuits. Integrated circuit chips and boards require not only high density but high reliability connections to other integrated circuit chips and boards to facilitate the manufacture of more complex products. The higher circuit density mandates the higher connection reliability because the probability of the product **fail**ing increases with the rising number of circuit connections within that product. Therefore, in order to maintain the product reliability, the reliability of the individual connections must increase. In addition, the connections must maintain this reliability even though the connections are connected, separated, and reconnected several times during the manufacturing of the product. This connecting and reconnecting process is necessary to test the individual electronic circuit components before they are assembled into a final product.*

Interpretation of interaction: Two important design attributes of electronic chips is circuit density (one function - for faster processing) and reliability (another function). However, when circuit density increases (improves), the reliability decreases, thus **these two functions interact**.

23. |3DPrinting|US5945058|Method and apparatus for identifying surface features associated with selected lamina of a three-dimensional object being stereolithographically formed|

*Because there is typically no process to form an initial coating of excess thickness (e.g. deep dipping where the object is dipped into the building material by more than one layer thickness) in a typical layer formation process using the applicator blade (i.e. a material dispensing device), not only must the applicator supply resin over the regions solidified in association with the formation of the last lamina but it must also add resin into the trapped volume as it moves across that region. If the applicator blade **fail**s to provide exactly the right amount of resin into the trapped volume region, any excess liquid may not flow from the trapped volume region quickly enough or any shortage of resin might not be filled in quickly enough due to the flow restrictions inherent therein. Though, these problems may be minor when considering the formation of a single layer, they can be problematic, if not catastrophic, to part building when a plurality of adjacent layers continue the formation of a trapped volume, whereby these errors can be accumulated. Consequently, trapped volume regions are problematic to defining generalized optimized recoating styles when using either a doctor blade or an applicator blade.*

Interpretation of interaction: The trapped volume regions are a **harmful side effect** of the process being described, in the sense that failure to fill in or remove excess from the trapped volume regions leads to errors in building the part.

24. |Magnetic Storage|US5668679|System for self-servowriting a disk drive|

*Conventional servowriters are expensive. Typically, a drive to be servowritten must be servowritten with its cover removed or with at least two external openings to permit the insertion of the clock head and the arm positioning mechanism when the drive is mounted on the servowriter. The requisite openings or holes give rise to several problems because the holes adversely affect the mechanical integrity of the disk drive's base casting or baseplate, thus causing stiffness and resonance related concerns. Also, the cutouts provide an ingress for contaminants which can easily ruin a disk drive. Servowriting with the drive's cover removed amplifies these problems. The entry of contaminants not only increases tribology related problems, but more importantly, because the spacing between a head and disk is typically not more than a few microinches, the entry of even microscopic particulate contaminates can lead to a catastrophic **fail**ure such as a head-to-disk crash.*

Interpretation of interaction: The openings in servodrives provide access (desired function); however, it also creates **harmful side effects**: allows ingress of contaminants, reduction of mechanical integrity and susceptibility to vibration.

25. |Integrated Circuits|US6236059|Memory cell incorporating a chalcogenide element and method of making same|

*Typically, a phase-change memory device includes a polysilicon lower electrode, also known as a matchstick. One challenge of forming a lower electrode in a phase-change memory cell is to shrink the cell size while still being able to dope the polysilicon matchstick structure in an ever-increasing aspect ratio recess. As the aspect ratio of the recess increases, it becomes increasingly difficult to properly dope the matchstick structure for at least two reasons. First, an increasingly steep angle of implantation directed at the polysilicon wall will result in an increasingly higher incidence of ricochet of the dopants instead of implantation. Second, as that aspect ratio gets higher, it becomes increasingly difficult to get dopant to strike the polysilicon wall at the bottom of the recess; an inadequate doping at the bottom results in a conductive **fail**ure.*

Interpretation of interaction: It is advantageous to reduce size (i.e. increase density to improve information storage – one function). However, reduction in size aggravates the ease of doping the polysilicon matchstick (manufacturability - second function). These **two functions are interacting**, where improvement of the first leads to reduction of the second.

26. |Superconductors|US4589001|Quasiparticle injection control type superconducting device|

*In the conventional cryogenic devices of this nature, the commonest method adopted for controlling the characteristics of the Josephson junction element comprises disposing control lines close to the element and effecting the required control using the magnetic fields generated by the control lines. By this method, however, the number of layers required for the lines inevitably increases, raising the chance of the devices suffering from line breakage, short-circuiting and other **fail**ures and from degradation of reliability of performance.*

Interpretation of interaction: The presence of control lines allows improvement of regulation (improves stability of performance of the Josephson element - one function). As number of control lines are increased for better control, the increase, however, leads to reduction in reliability due as high chance of short-circuiting and other failures. The **two functions - performance and reliabil**ity - of the element **interact** in this case.

27. |LED|US4720432|Electroluminescent device with organic luminescent medium|

*While the aromatic tertiary amine layers employed by Van Slyke et al, cited above, have resulted in highly attractive initial light outputs in organic EL devices, the limited stability of devices containing these layers has remained a deterrent to widespread use. Device degradation result in obtaining progressively lower current densities when a constant voltage is applied. Lower current densities in turn result in lower levels of light output. With a constant applied voltage, practical EL device use terminates when light emission levels drop below acceptable levels--e.g., readily visually detectable emission levels in ambient lighting. If*

*the applied voltage is progressively increased to hold light emission levels constant, the field across the EL device is correspondingly increased. Eventually a voltage level is required that cannot be conveniently supplied by the EL device driving circuitry or which produces a field gradient (volts/cm) exceeding the dielectric breakdown strength of the layers separating the electrodes, resulting in a catastrophic **fail**ure of the EL device.*

Interpretation of interaction: The progressive increase in voltage provides stable light output (performance – one function). However, as the voltage is increased to maintain light, it eventually leads to a **harmful side effect**, in which the dielectric breaks down causing catastrophic failure of the device.

28. |Flywheel|US7263912|Flywheel hub-to-rim coupling|

*The flywheel system is normally contained in a vacuum enclosure that protects it from windage losses that would occur from operation in a gas atmosphere, and provides ballistic protection against catastrophic **fail**ure of a flywheel rotating at high speed.*

Interpretation of interaction: The **text does not indicate interaction**.

**Keyword: Disadvantage**

29. |Combustion engine|US4856465|Multidependent valve timing overlap control for the cylinders of an internal combustion engine|

*The control of intake and exhaust valves by means of a cam shaft in Otto engines of conventional construction involves a compromise regarding the overlap of the intake and exhaust valving generally known as the valve overlap Vo. As the result of this compromise, at small speeds and loads a small valve overlap is provided for reasons of quietness of operation and exhaust gas quality, among others, while at high engine speeds and loads, on the other hand, a large valve overlap is used because of advantages in obtaining high power development. Valving control in accordance with present practice, applied to a cam shaft for*

*control of the intake valve, typically keeps Vo small below a switchover rotary speed of about 1600 r.p.m. Because of the low level of internal exhaust gas recycling, the running quietness of the engine is good and the hydrocarbon emissions are low (especially in idling and at low load). About the switchover speed, a high value of Vo is mechanically set so that at high load a high torque will be generated. A **disadvantage** of the known system is, on the one hand, the high engineering and manufacturing cost of perfecting and producing mechanically speed-controlled pressure release valves that switch over only at a quite precisely determined shaft speed. Variable switchover depending upon speed, particularly in accordance with a defined function of speed, is not possible in such arrangements. Furthermore, the cam shaft must be expensively disassembled in the case of malfunction or damage of the centrifugal hydraulic switch. It has further been found that at low speed and high load (e.g., acceleration phases) the nitrogen oxide emissions are greater and a lower torque is developed than would be the case with a higher Vo.*

> Interpretation of interaction: The valving overlap provides quiet operation and better quality exhaust at low speeds and provides high power at high speeds (one function). However, this overlapping approach leads to increase in engineering and manufacturing cost (another function) and reduction in ease of repairing (still another function). Thus in this compromise there is **interaction between functions**.

30. |Fuel Cell|US5364711|Porous fuel electrode, electrolyte plates, oxidizing electrode, fuel diffusion means, oxidizing gas; miniaturization|

*As mentioned in the relevant specification, this fuel cell has a constructional limitation that the liquid fuel is capable of permeating or penetrating the anode (fuel electrode) in the horizontal direction but is incapable of succumbing to the capillary action in the upward direction. Further, since the fuel cell of this system is so constructed as to necessitate interposition of a gap between the bottom surface of the stack and the fuel storing chamber and insertion of the fibrous capillary material in the fuel storing chamber as described above, the capillary material and the fuel storing chamber allow no easy sealing and, at the same time, the stack and the fuel storing chamber must be so constructed as to be integrally fixed. Moreover, since the fuel storing chamber must be provided in the upper part thereof with a*

*plurality of inlets for the introduction of the capillary material in preparation for the integration, it entails the **disadvantag**e that the construction thereof is complicated and extremely difficult to manufacture. It may be safely added that the construction for this integral fixation, the necessary for opening slits at least in the oxidizing electrode part of the bottom surface of the stack for the purpose of ensuring the flow of the oxidizing agent also adds to the complication of the construction. Since the liquid fuel is supplied by the capillary action which is manifested in one direction from the lower part to the upper part as described above, the travel of the fuel to the upper part of the fuel electrode consumes much time and the shape of the fuel cell places a limit on the supply of the fuel by the capillary action.*

Interpretation of interaction: The multiple inlets with fibrous material allow flow of fuel using capillary action required for the fuel cell; however, the multiple inlets reduce the manufacturability of the device. Thus, **two functions - performance (energy conversion) and manufacturing (another function) - are interacting**.

31. |Incandescent lighting|US6555948|Electric incandescent lamp|

*The already mentioned EP 0 588 541 has therefore addressed the object of proposing an electric incandescent lamp in which the helix and the layer reflecting IR radiation are arranged relative to one another in an essentially unfocussed relationship, and yet satisfactory absorption of IR radiation is ensured. In order to achieve this object, EP 0 588 541 provides an incandescent filament which comprises coiled segments of tungsten wire which are connected to one another and are supported by segment bearings in between the segments in an essentially rectangular frame. … A substantial **disadvantag**e exists, in particular, in that the coiling of the tungsten wire leads to a so-called "radiation blackening". To be specific, because of the temperature-dependence of its spectral emission coefficient, pure tungsten, which is preferably applied as filament material, has a light yield which is higher at the same temperature by circa 40% than the black body. This gain in selectivity is lost in part upon coiling the wire. It would be possible to counter a reduction in the radiation blackening by enlarging the pitch. However, this would contradict the requirement for compact filaments.*

Interpretation of interaction: The coiling of the tungsten wire helps to reduce IR radiation losses (one function); coiling, however, causes radiation blackening, a **harmful side effect** which leads to reduction in light emission.

32. |Electric motor|US5117141|Disc rotors with permanent magnets for brushless DC motor|

*It will be appreciated that conventional DC motors that employ slip rings and/or brushes, suffer a number of **disadvantag**es and problems, including dirt and dust buildup associated with the slip rings, brushes, gearbox speed reducers and the like. Further, the arcing associated with such slip rings and brushes can create a serious hazard in underwater vehicles.*

Interpretation of interaction: The disadvantages described are associated **harmful side effects** – arcing, and dust build up.

33. |Electronic computation|US6732068|Memory circuit for use in hardware emulation system|

*However, altering or re-synthesizing the clock paths in an asynchronous design can lead to inaccurate or misleading emulation results. Since most circuit designs have asynchronous clock architectures, the need to alter or re-synthesize the clock paths is a large **disadvantag**e. In addition, prior art hardware emulation machines using time-multiplexing have suffered from low operating speed.*

Interpretation of interaction: The re-synthesis of clock paths in asynchronous design produce inaccurate or misleading emulation results, which are **harmful side effects** that need to be dealt with.

34. |Superconductors|US4965245|Method of producing oxide superconducting cables and coils using copper alloy filament precursors|

*These methods are **disadvantag**eous in that it is difficult to produce long superconducting wires or cables with excellent superconductivity. The reason is that it is rather hard to completely uniformly mix superconductor material powders with the result in that it is difficult to obtain a homogeneous crystal structure of the superconductor along the entire length of the cable.*

Interpretation of interaction: The inherent difficulty in mixing powers leads to non-uniformity in the superconducting wires, a **harmful side effect** that limits the length of superconductors that can be produced.

35. |Semiconductor storage|US6859389|Phase change-type memory element and process for producing the same|

*The structure of the phase change-type memory element disclosed in Japanese Patent Laid-Open No. 21740/1993 is such that a recording material (a phase change material) is provided between electrodes, and an insulating layer with an opening having a predetermined size is interposed between the electrodes and the recording material. Upon the application of a set pulse across the electrodes, the phase state is brought to an ON state, and upon the application of a reset pulse, the phase state is returned to an OFF state. Since, however, the diameter of a current path formed in the recording material upon the application of voltage is as large as 2 to 3 m, the volume of the phase change area is large. This poses a problem that a large current pulse is necessary as a reset pulse for returning the phase state from the ON state to the OFF state. Further, the recording material except for the phase change area should be in an amorphous state. This necessitates the production of the phase change-type memory element at a temperature at or below the crystallization temperature of the recording material. Therefore, for example, the temperature in the production of transistors or diodes constituting a drive circuit is **disadvantag**eously restricted.*

Interpretation of interaction: The recording material undergoes phase change in a much wider area than required. The alteration of unnecessary area in the recording material is a **harmful side effect**, as it makes it necessary to spend more energy to reset the recording medium back to the amorphous state, as well as, to operate in a more restricted temperature conditions, thus reducing the ease of manufacturing.

**Keyword: Overcome**

36. |Magnetic Storage|US4135217|Utilization of stored run-out information in a track following servo system|

*Wobble is caused by a number of factors including, for example, 1) the free play or fit of the shaft, rotating the disk, on bearings, 2) the stiffness of the bearings, and 3) the mounting of the disks on the shaft, which can cause eccentric shaft rotation. Wobble can be reduced, but this requires more expensive bearings to increase their stiffness, and more expensive mountings to reduce free play and provide a truer circular shaft rotation. For example, there has been used conically shaped shafts and mating hubs to provide a tighter, more accurate fit. If the bandpass of the servo system is increased, then the above-mentioned error for a disk rotating at 40 Hz, can be reduced. However, as with the solution to reducing wobble, this requires more expensive components. The system bandpass can be increased but only at the expense of using more massive parts, such as heavier carriages and carriage arms supporting the heads. Moreover, the more massive carriage and arms supporting the heads would require greater power consumption by the servo system electrical components to move them.|"It is an object of the present invention to provide a novel apparatus for positioning one member with respect to another member. It is another object of the present invention to provide a servo system which **overcom**es the above-mentioned disadvantages of prior solutions for improving the accuracy of positioning a recording head in relation to a disk having high density tracks of data.*

Interpretation of interaction: The wobble is a **harmful side effect** which reduces the accuracy of position of recording heads. The text describes some potential counter-measures to overcome the wobble.

37. |Electronic computation|US6691301|System, method and article of manufacture for signal constructs in a programming language capable of programming hardware architectures|

*However, for a given function, a software-controlled processor is usually slower than hardware dedicated to that function. A way of **overcom**ing this problem is to use a special software-controlled processor such as a RISC processor which can be made to function more quickly for limited purposes by having its parameters (for instance size, instruction set etc.) tailored to the desired functionality. Where hardware is used, though, although it increases the speed of operation, it lacks flexibility and, for instance, although it may be suitable for the task for which it was designed it may not be suitable for a modified version of that task which is desired later. It is now possible to form the hardware on reconfigurable logic circuits, such as Field Programmable Gate Arrays (FPGA's) which are logic circuits which can be repeatedly reconfigured in different ways. Thus they provide the speed advantages of dedicated hardware, with some degree of flexibility for later updating or multiple functionality. In general, though, it can be seen that designers face a problem in finding the right balance between speed and generality.*

Interpretation of interaction: The dedicated software-controlled processors increase speed (one function) but at the expense of sacrificing flexibility to handle other types of applications (another function). Therefore, the **two functions - performance (speed) and flexibility - are interacting**.

38. |Optical Telcom|US4244045|Optical multiplexer and demultiplexer|

*Said U.S. patent also discloses a multiplexer in which a plurality of band pass filters are arranged around a glass plate with semi-reflective walls. However, this multiplexer has the disadvantage that the loss of the light beam is great since the light beam suffers from a plurality of partial reflections or partial transmission in said semi-reflective walls."| However, this multiplexer has the disadvantage that the loss of the light beam is great since the light beam suffers from a plurality of partial reflections or partial transmission in said semi-*

*reflective walls."|It is an object, therefore, of the present invention to **overcom**e the disadvantages and limitations of prior multiplexer and/or demultiplexer by providing a new and improved optical-multiplexer and/or demultiplexer.*

Interpretation of interaction: The reflection from successive mirrors is utilized for multiplexing and demultiplexing the signal. The transmitted beam suffers from a **harmful side effect** of losing intensity of the beam (or weakening the signal) as the beam undergoes multiple reflections.

39. |Optical Storage|US5276662|Disc drive with improved data transfer management apparatus|

*Since the system microprocessor provides overall control of the disc drive, the time required for executing these interrupt routines can limit the capabilities of the disc drive solely on the basis that the microprocessor processing time is not available to carry out functions required to provide additional capabilities. Thus, while the general scheme for effecting data transfers between a host and a disc is, in principle, an effective approach to data storage and retrieval, the implementation of the scheme has given rise to practical problems that have not heretofore been solved." |The present invention **overcom**es these practical problems by including in a disc drive a hardware buffer manager that effects sector level control of data transfers.*

Interpretation of interaction: The data transfer from the disc is controlled by the system microprocessor. This is good from the viewpoint of having fewer components. However, availability of the microprocessor for executing routines is limited, hence it cannot service the disc all the time. This limits the functionality of the disc (another function). Here **two functions are interacting** – cost and speed of storage retrieval.

40. |Electric Telcom|US6331123|Connector for hard-line coaxial cable|

*The radial crimping, however, often does not apply compressive force evenly to the outer conductor or alternatively to outer tubular jacket of the outer connector. Such uneven compression can form channels for infiltration of moisture into the coaxial cable connection and consequently leading to the degradation of the signal carried by the cable. … To **overcom**e the difficulties of crimped and threaded connectors different designs of axially compressible connectors have been employed.*

Interpretation of interaction: The non-uniformity of compressive force in radial crimping leads to a **harmful side effect** of moisture infiltration which degrades the signal.

41. |Superconductors|US5350738|Method of manufacturing an oxide superconductor film|

*… in the method (2) above, film formation to a region of a large area can be conducted relatively easily since there is less restriction in view of the constitution of the device. However, this method inevitably suffers from intrusion of impurities upon vapor deposition due to residual gases such as $H_2O$, $N_2$, $H_2$ and $CO_2$. Further, since $BaF_2$ is used as an evaporation source, it is necessary to remove a F-content in the film upon oxidation and crystallization by reaction with steams at a high temperature in addition to reaction with oxygen. As a result, this not only makes the operation complicate but also generates a toxic HF gas of highly corrosive nature during reaction to bring about a problem in view of safety and sanitation. Further, since the evaporation state of $BaF_2$ is instable, it is difficult for the accurate control of the film composition. The present invention has been achieved in order to **overcom**e the technical problems involved in the prior art and it is an object thereof to provide a method of manufacturing an oxide superconductor film capable of easily controlling a film-forming rate and a film composition, producing the film safely and economically over a wide region, homogeneously and at a high quality.*

Interpretation of interaction: The method described allows easy formation of a film. However, it suffers from number of **harmful side effects** such as formation toxic HF gas.

42. |3DPrinting|US6730256|Stereolithographic patterning with interlayer surface modifications|

*Lastly, solvent weakening has been attempted in only a few limited cases. Different materials (or resists with different molecular weights) are used for each layer. The processing conditions used to apply the bottom layer either would destroy the photoimaging properties of the top layer preventing continued stacking of this composite structure or the process is only compatible with a single exposure for both layers. A need exists for a method that **overcom**es the deficiencies of the aforementioned stereolithographic methods. A technique that leverages advancements in high resolution photoresists and coating techniques already developed for the patterning of integrated circuits without the extra complexity of intermediate plating and development steps would satisfy a long-felt need in the art.*

Interpretation of interaction: The method described suffers from a **harmful side effect** in which the photoimaging properties of the top layer is destroyed, which prevents continued stacking.